%% file: UEFA_match_fixing_qualification_draw_v2.tex
\documentclass[12pt]{article}
\usepackage[latin1]{inputenc}
\usepackage[british]{babel}
\usepackage{cmap}
\usepackage{lmodern}

\usepackage{amssymb, amsmath, amsthm}
\usepackage[a4paper,top=25mm,bottom=25mm,left=25mm,right=25mm]{geometry}
\usepackage{ragged2e}

\usepackage{authblk} 
\usepackage{pifont}
\usepackage{graphicx}
\usepackage[usenames,dvipsnames,svgnames,table]{xcolor}
\usepackage[figuresright]{rotating}
\usepackage{xtab} 
\usepackage{longtable} 
\usepackage{multirow}
\usepackage{footnote}
\usepackage[stable]{footmisc}
\usepackage{chngpage} 
\usepackage{pdflscape} 
\usepackage[nottoc,notlot,notlof]{tocbibind} 
\usepackage{bookmark} %

\usepackage{pgfplots}
\pgfplotsset{compat=1.14}
\pgfplotsset{every tick label/.append style={font=\footnotesize}}
\usepackage{setspace}

\makesavenoteenv{tabular}
\usepackage{tabularx}
\usepackage{booktabs}
\usepackage{threeparttable} 
\usepackage[referable]{threeparttablex} 
\newcolumntype{R}{>{\raggedleft\arraybackslash}X}
\newcolumntype{L}{>{\raggedright\arraybackslash}X}
\newcolumntype{C}{>{\centering\arraybackslash}X}
\newcolumntype{A}{>{\columncolor{gray!25}}C}
\newcolumntype{a}{>{\columncolor{gray!25}}c}

\newlength{\tablen}

\usepackage{dcolumn} 
\newcolumntype{.}{D{.}{.}{-1}}

\usepackage{tikz}
\usetikzlibrary{arrows, calc, matrix, patterns, positioning, shapes, trees}
\usepackage[semicolon]{natbib}
\usepackage{hyperref} 
\hypersetup{
  colorlinks   = true,    
  urlcolor     = blue,    
  linkcolor    = blue,    
  citecolor    = ForestGreen      
}
\usepackage{microtype}
\usepackage[singlelinecheck=false,justification=centering]{caption} 

\usepackage[labelformat=simple]{subcaption}

\DeclareCaptionLabelFormat{parenthesis}{(#2)}
\makeatletter
\renewcommand\p@subfigure{\arabic{figure}.}
\makeatother

\DeclareCaptionLabelFormat{parenthesis}{(#2)}
\makeatletter
\renewcommand\p@subtable{\arabic{table}.}
\makeatother

\usepackage{enumitem}
\setlist[itemize]{leftmargin=2.5\parindent}
\setlist[enumerate]{leftmargin=2.5\parindent}

\def\addlegendimage{\csname pgfplots@addlegendimage\endcsname}

\theoremstyle{plain}

\theoremstyle{definition}

\newtheorem{example}{Example}

\theoremstyle{remark}

\def\keywords{\vspace{.5em} 
{\noindent \textit{Keywords}: }}

\def\JEL{\vspace{.5em} 
{\noindent \textbf{\emph{JEL} classification number}: }}

\def\AMS{\vspace{.5em} 
{\noindent \textbf{\emph{MSC} class}: }}

\author{\href{https://sites.google.com/view/laszlocsato}{L\'aszl\'o Csat\'o}\thanks{~E-mail: \emph{laszlo.csato@sztaki.hu}} }
\affil{Institute for Computer Science and Control (SZTAKI) \\
E\"otv\"os Lor\'and Research Network (ELKH) \\
Laboratory on Engineering and Management Intelligence \\
Research Group of Operations Research and Decision Systems}
\affil{Corvinus University of Budapest (BCE) \\
Department of Operations Research and Actuarial Sciences}
\affil{Budapest, Hungary}
\title{Quantifying incentive (in)compatibility: \\ A case study from sports}
\date{\today}

\def\Dedication{
{\noindent
$\mathfrak{In}$ $\mathfrak{der}$ $\mathfrak{Regel}$ $\mathfrak{ist}$ $\mathfrak{jeder}$ $\mathfrak{geneigt}$, $\mathfrak{das}$ $\mathfrak{Schlimme}$ $\mathfrak{eher}$ $\mathfrak{zu}$ $\mathfrak{glauben}$ $\mathfrak{als}$ $\mathfrak{das}$ $\mathfrak{Gute}$; $\mathfrak{jeder}$ $\mathfrak{ist}$ $\mathfrak{geneigt}$, $\mathfrak{das}$ $\mathfrak{Schlimme}$ $\mathfrak{etwas}$ $\mathfrak{zu}$ $\mathfrak{vergr\ddot{o}\ss ern}$, $\mathfrak{und}$ $\mathfrak{die}$ $\mathfrak{Gef\ddot{a}hrlichkeiten}$, $\mathfrak{welche}$ $\mathfrak{auf}$ $\mathfrak{diese}$ $\mathfrak{Weise}$ $\mathfrak{berichtet}$ $\mathfrak{werden}$, $\mathfrak{ob}$ $\mathfrak{sie}$ $\mathfrak{gleich}$ $\mathfrak{wie}$ $\mathfrak{die}$ $\mathfrak{Wellen}$ $\mathfrak{des}$ $\mathfrak{Meeres}$ $\mathfrak{in}$ $\mathfrak{sich}$ $\mathfrak{selbst}$ $\mathfrak{zusammensinken}$, $\mathfrak{kehren}$ $\mathfrak{doch}$ $\mathfrak{wie}$ $\mathfrak{jene}$ $\mathfrak{ohne}$ $\mathfrak{sichtbare}$ $\mathfrak{Veranlassung}$ $\mathfrak{immer}$ $\mathfrak{von}$ $\mathfrak{neuem}$ $\mathfrak{zur\ddot{u}ck}$.\footnote{~
``\emph{As a general rule every one is more inclined to lend credence to the bad than the good. Every one is inclined to magnify the bad in some measure, and although the alarms which are thus propagated, like the waves of the sea, subside into themselves, still, like them, without any apparent cause they rise again.}'' (Source: Carl von Clausewitz: \emph{On War}, Book 1, Chapter 6 [Information in War]. Translated by Colonel James John Graham, London, N. Tr\"ubner, 1873. \url{http://clausewitz.com/readings/OnWar1873/TOC.htm})}
}
\vspace{0.25cm}

\flushright
\noindent (Carl von Clausewitz: \emph{Vom Kriege})

\vspace{1cm} 
\justify }

\begin{document}

\newgeometry{top=5mm,bottom=15mm,left=25mm,right=25mm}

\maketitle
\thispagestyle{empty}
\Dedication

\begin{abstract}
\noindent
For every sports tournament, it is an important requirement to provide contestants with the appropriate incentives to perform. However, incentive compatibility is usually considered an all or nothing (binary) concept in the academic literature, that is, the rules are proved to be either strategy-proof or vulnerable to manipulation.
Our paper aims to present a method for quantifying the violation of this theoretical property through the example of the European Qualifiers for the 2022 FIFA World Cup. Even though that competition is known to be incentive incompatible since lower-ranked Nations League group winners are interested in the success of their higher-ranked peers, the extent of the problem has remained unexplored until now. Computer simulations reveal that the threat of tanking can be substantially mitigated by adding a carefully chosen set of draw restrictions, which offers a justifiable and transparent solution to improve fairness. Sports governing bodies are encouraged to take our findings into account.

\keywords{OR in sports; draw procedure; FIFA World Cup; incentive compatibility; simulation}

\AMS{62F07, 90-10, 90B90, 91B14}

\JEL{C44, C63, Z20}
\end{abstract}

\clearpage
\restoregeometry

\section{Introduction} \label{Sec1}

Tournaments are extensively used as a tool to elicit costly effort from economic agents. According to recent theoretical results \citep{Pauly2014, Vong2017, DagaevSonin2018}, sports tournaments are usually not immune to strategic manipulation by the players, which is detrimental to the reputation of the organisers and threatens the profitability of the industry in the long run. Nonetheless, the trade-off between incentive compatibility and other desirable properties remains largely unexplored. A severe impediment is that strategy-proofness is traditionally regarded as an all or nothing (binary) concept and no attempt has ever been made to quantify the level of its violation in the tournament design literature.

The present paper aims to fill this research gap. Since it is close to impossible to devise a general abstract model for measuring the lack of strategy-proofness due to the huge variation of incentive schemes applied in the real world, we focus on group-based qualification systems with an exogenous ranking of the contestants used to provide a secondary way to qualify. Consequently, the lower-ranked contestants are interested in the success of higher-ranked contestants to fill the vacancies created by the latter, this design violates incentive compatibility \citep[Proposition~3]{DagaevSonin2018}, a basic principle of competitive sport \citep{Szymanski2003}.

However, the Union of European Football Association (UEFA) designs its qualification tournaments for the FIFA World Cup and the UEFA European Championship according to that principle since launching the UEFA Nations League in 2018. Therefore, it is crucial to avoid games with misaligned incentives to the maximum extent possible because no team can prove to play honestly, hence even the mere existence of such a situation endangers the integrity of the competition in the mind of spectators.

Fortunately, the lack of an incentive compatible solution does not exclude the possibility that the relevance of the problem can be reduced by a well-devised mechanism. In fact, we uncover how adding an appropriate set of restrictions in the group stage draw is able to significantly improve fairness. The intuition behind the idea is straightforward: two teams assigned to different groups will not play any match, thus none of them can benefit from \emph{tanking} (deliberately losing) against the other.

In particular, the European Qualifiers for the 2022 FIFA World Cup will be analysed as a case study, where the probability of an incentive incompatible scenario will be determined. Even though the violation of this theoretical property is based on an exact mathematical result, analytical tools remain insufficient to measure the extent of the deficiency: since only some teams have misaligned incentives in certain games, the first step would be to estimate the probability of these matches but even the group draw procedure is prohibitively complex \citep{Csato2021m}, not to say anything about the prediction of match outcomes.
Therefore, we are forced to use Monte Carlo simulations, a standard approach in the comparison and evaluation of tournament designs.

The central contribution of the current work resides in quantifying the incentive compatibility of a sports tournament that has never been attempted before. The methodology can be followed in the case of other incentive incompatible formats such as when teams playing in different groups should be compared \citep{Csato2020c}.
It is also shown first how additional draw constraints might be used to improve fairness. Finally, our proposal essentially solves the main problem of this particular tournament design, which can promote its application in the future, especially because it does not require any fundamental rule change, only the introduction of further prohibited clashes that can be justified by competition-related reasons.

The paper is structured as follows.
Section~\ref{Sec2} gives a concise overview of the literature. The format of the European qualifiers for the 2022 FIFA World Cup is outlined in Section~\ref{Sec3}. Our approach to quantifying incentive incompatibility is detailed in Section~\ref{Sec4}. Section~\ref{Sec5} reports the computational results and justifies the introduction of additional draw constraints. Section~\ref{Sec6} concludes.

\section{Related literature} \label{Sec2}

Cheating in sports is an important topic of scientific research because this behaviour violates most ethical codes and can potentially undermine interest. \citet{PrestonSzymanski2003} discuss different forms of cheating, including sabotage, doping, and match fixing. \citet{KendallLenten2017} present sports rules with unforeseen consequences, while \citet{BreuerForrest2018} offer a wide coverage of the most serious manipulations in sports. \citet{Andreff2019} overviews the economic aspects of criminal behaviour in sports.

There are also instructive case studies. According to \citet{DugganLevitt2002}, sumo wrestlers coordinate their fights when approaching eight victories towards the end of the season due to the sharp non-linearity in the payoff function. In football games where one team is in immediate danger of relegation to a lower division but the other team is not affected by the outcome, the former club achieves a desired result with a higher probability in more corrupt countries \citep{ElaadKrumerKantor2018}. As recent regression results show, bookmakers and bettors believe that teams do not exert to the maximum extent to win once they have no chance of qualifying for international cups or being relegated \citep{FeddersenHumphreysSoebbing2021}.


Perhaps the most famous example of dual incentives is caused by the player draft in North American (and Australian) sports because the traditional set-up of reverse order promotes tanking once a team has been eliminated from the play-offs. \citet{TaylorTrogdon2002} examine three National Basketball Association (NBA) seasons to determine whether performance responds to changes in the lottery system used for the draft order. Teams are found to be more likely to lose in the presence of incentives to lose. \citet{PriceSoebbingBerriHumphreys2010} reinforce these results on a larger dataset. While losing more games after elimination can merely be a consequence of lower motivation and disappointment, this is not the only explanation: a concrete strategy behind losing can be identified, at least in the National Hockey League (NHL) \citep{Fornwagner2019}. Consequently, several remedies have been proposed to adjust the draft allocation mechanism in order to mitigate the threat of tanking \citep{BanchioMunro2021, Gold2010, KazachkovVardi2020, Lenten2016, LentenSmithBoys2018}---and they have been partially adopted by the administrators as shown by the continuous development of the \href{https://en.wikipedia.org/wiki/NBA_draft_lottery}{NBA draft lottery}.

\citet{Pauly2014} proves an impossibility theorem to show that complex tournament systems consisting of two qualifying tournaments with disjoint sets of participants, such as the design of recent FIFA World Cups, give rise to manipulation.
According to \citet{Vong2017}, the necessary and sufficient condition of strategy-proofness in multistage tournaments is strongly restrictive since only the top-ranked player can be allowed to qualify from each group.
\citet{DagaevSonin2018} consider incentive compatibility in multiple tournaments with the same set of participants and noncumulative prizes when the sets of winners have a non-empty intersection. Their main theorem essentially implies that the tournament examined in the present paper can be manipulated. The ignorance of this theoretical result is responsible for the misaligned incentive schemes in the UEFA Champions League entry between the 2015/16 and the 2017/18 seasons \citep{Csato2019c}, as well as in the current Champions League group stage draw \citep{Csato2020a}. \citet{Csato2020c} analyses the incentive compatibility of group-based qualification systems when teams from different groups are compared to provide a secondary chance for qualification, where playing a draw can be optimal for both teams under certain circumstances \citep{Csato2020d}.

However, all of the previous works treat incentive (in)compatibility as an all or nothing (binary) concept and do not undertake to quantify the threat of manipulation. Naturally, strategy-proofness is extensively discussed in other fields of research but we know only a few attempts to measure the seriousness of its violation. \citet{AltmanTennenholtz2006} evaluate the incentive compatibility of ranking systems. \citet{BonkoungouNesterov2021} investigate the criterion of strategic accessibility for school choice mechanisms, that is, the set of schools wherein a student can get admission by manipulation.

On the other hand, it is a standard approach in scheduling to maximise the competitiveness of the matches played in the last round of round-robin tournaments \citep{ChaterArrondelGayantLaslier2021, Guyon2020a, Stronka2020}. Similarly, \citet{LasekSzlavikGagolewskiBhulai2016} assess opportunities for advancing in the previous FIFA World Ranking table in terms of associated probabilities.
While it may seem strange that quantifying the level of incentive incompatibility in tournament design has never been the subject of any scientific study, a possible explanation can be the complicated structure of real-world competitions. 

Finally, since we will recommend adding further draw constraints, the paper is strongly connected to studies on the group draw in sports tournaments, too. \citet{Guyon2015a} identifies several flaws in the draw of the 2014 FIFA World Cup and provides a fair and evenly distributed method to create balanced and geographically diverse groups. \citet{LalienaLopez2019} and \citet{CeaDuranGuajardoSureSiebertZamorano2020} propose alternative policies to balance the difficulty levels of the groups in the FIFA World Cup.  In the case of the UEFA Champions League knockout stage draw, the focus is on the fairness of the draw procedure that can be improved by relaxing the restrictions \citep{KlossnerBecker2013, BoczonWilson2018}. But this is the first study to consider how draw conditions can be used to achieve other desirable goals than the equal treatment of all teams.

\section{A deficiency of the European Qualifiers for the 2022 FIFA World Cup} \label{Sec3}

The European section of the 2022 FIFA World Cup qualification is a football competition contested by the national teams of the 55 UEFA member associations to allocate the 13 slots available for this confederation in the World Cup finals.
The tournament consists of two phases. In the first group stage, the countries are put into ten groups with five or six teams each to play in a home-away round-robin structure. The group winners qualify directly for the 2022 FIFA World Cup, while the runners-up advance to the second play-off stage. Here, the two highest-ranked group winners from the 2020/21 UEFA Nations League that finished outside the top two in their groups join them. These 12 teams are divided randomly into three play-off paths of four teams each, and the three path winners qualify for the World Cup finals.

According to the model of \citet{DagaevSonin2018}, the above design is vulnerable to strategic manipulation: if the Nations League ranking is $A \succ B \succ C$ such that all teams play in the same round-robin group of the qualifiers and teams $A$ and $C$ compete for the prize, team $B$ has to lose against team $A$ in order to guarantee the qualification of team $A$ ahead of team $C$ because then team $B$ could have a chance to fill the vacant slot in the play-offs. \citet{HaugenKrumer2021} identify such a situation in the UEFA Euro 2020 qualifying concerning Israel and Austria.

\begin{table}[t!]
  \centering
  \caption{The ranking of group winners in the 2020/21 UEFA Nations League}
  \label{Table1}
    \rowcolors{1}{}{gray!20}
\begin{threeparttable}
    \begin{tabularx}{0.8\textwidth}{l Ll cc} \toprule
    League & Group winner & Abbreviation & Rank  & Place in the draw \\ \bottomrule
    A     & France & FRA   & 1     & Pot 1 \\
    A     & Belgium & BEL   & 2     & Pot 1 \\
    A     & Italy & ITA   & 3     & Pot 1 \\
    A     & Spain & ESP   & 4     & Pot 1 \\ \hline
    B     & Wales & WAL   & 17    & Pot 2 \\
    B     & Austria & AUT   & 18    & Pot 2 \\
    B     & Czech Republic & CZE   & 19    & Pot 3 \\
    B     & Hungary & HUN   & 20    & Pot 3 \\ \hline
    C     & Slovenia & SVN   & 33    & Pot 4 \\
    C     & Montenegro & MNE   & 34    & Pot 4 \\
    C     & Albania & ALB   & 35    & Pot 4 \\
    C     & Armenia & ARM   & 36    & Pot 5 \\ \hline
    D     & Gibraltar & GIB   & 49    & Pot 6 \\
    D     & Faroe Islands & FRO   & 50    & Pot 5 \\ \bottomrule
    \end{tabularx}
\begin{tablenotes} \footnotesize
\item
The last column shows the pot of the team in the draw of the European Qualifiers for the 2022 FIFA World Cup.
\item
The first four spots are determined by the 2021 UEFA Nations League Finals to be played after the beginning of the qualifying tournament. These teams are ranked on the basis of their performance in 2020/21 UEFA Nations League A.
\end{tablenotes}
\end{threeparttable}
\end{table}

\begin{example} \label{Examp1}
Group I in the European Qualifiers for the 2022 FIFA World Cup contains England, Poland, Hungary, Albania, Andorra, and San Marino.
Consider the ranking of the 2020/21 UEFA Nations League group winners in Table~\ref{Table1}, which has been established on 18 November 2020, before the group draw of the qualifying tournament.
Albania has not a high chance to reach the first two positions in Group I. Nonetheless, it can still go to the play-offs if at least nine teams ranked higher than Albania in Table~\ref{Table1} obtain the first two spots in their groups. Consequently, a reasonable strategy for Albania would be to lose both of its group matches against Hungary, while exerting full effort against all other teams. This plan maximises the probability that Hungary finishes as the group winner or runner-up.
\end{example}

\citet[Chapter~6.2]{Csato2021a} offers a theoretical study of this tournament format to conclude that there is no obvious remedy of incentive incompatibility if UEFA wants to reward performance in the Nations League.
Therefore, we aim to mitigate the danger of a strategic manipulation outlined in Example~\ref{Examp1} to the maximum extent possible.

\section{Methodology} \label{Sec4}

The violation of incentive compatibility will be measured via Monte Carlo simulations, a standard approach in the tournament design literature \citep{ChaterArrondelGayantLaslier2021, Csato2020b, Csato2021b, DagaevRudyak2019, GoossensBelienSpieksma2012, ScarfYusofBilbao2009}.
Our starting point is the group stage draw of the European Qualifiers for the 2022 FIFA World Cup when the ranking of the Nations League group winners (Table~\ref{Table1}) has already been known.
The rules of the draw are as follows \citep{UEFA2020c}.

The 55 teams are seeded into six pots based on the November 2020 FIFA World Ranking such that the 10 strongest teams are placed in Pot 1, the next 10 in Pot 2, and so on until the five lowest-ranked countries are assigned to Pot 6. The teams are allocated to Groups A--E of five teams and Groups F--J of six teams. The countries are drawn sequentially from Pot 1 to Pot 6, and each team is allotted to the first available group in alphabetical order as indicated by the computer. Four types of draw constraints should be met to obtain an assignment ``\emph{that is fair for the participating teams, fulfils the expectations of commercial partners and ensures with a high degree of probability that the fixture can take place as scheduled}'' \citep{UEFA2020c}.
\citet{Csato2021m} discusses how the unfairness of the draw procedure can be reduced.

Once the group allocation is determined, every game can be simulated. The ability of a team is measured by a single variable, its Elo rating. Although there exist more developed prediction techniques \citep{BakerMcHale2018, CoronaForrestTenaWiper2019, LeyVandeWieleVanEeetvelde2019}, this relatively simple approach has been chosen because
(1) Elo rating is a good indicator of the real level of national football teams \citep{GasquezRoyuela2016};
(2) the decision-makers can better understand our central message;
(3) the main findings are unlikely to change in a more sophisticated parametric model.

Instead of the FIFA World Ranking, we use the Elo ratings of the popular project World Football Elo Ratings, available at \url{http://www.eloratings.net/}, as a proxy for team performances. Predictions from this Elo rating are more accurate than the previous FIFA ranking \citep{LasekSzlavikBhulai2013, GasquezRoyuela2016}, and---in contrast to the current FIFA World Ranking---it distinguishes home and away games as well as accounts for the margin of victory when the ratings are updated. In addition, although FIFA adopted the Elo formula in the summer of 2018 \citep{FIFA2018c}, the revision involved a smooth transition from the previous ranking, and the moderated number of matches played since then has probably not allowed yet for the necessary adjustments.

Take a match between teams $i$ and $j$ with the Elo ratings $E_i$ and $E_j$, respectively, played at the field of team $i$.
The \emph{win expectancy} of team $i$ is
\begin{equation} \label{Elo_equation}
W_{ij} = \frac{1}{1 + 10^{- \left( E_i + 100 - E_j \right) /400}}.
\end{equation}
Note that the rating of the home team is increased by 100 as travel fatigue, support from the crowd, referee bias, and familiarity with the stadium provide a significant home field advantage in football \citep{Pollard1986, BoykoBoykoBoyko2007, BakerMcHale2018}.

However, the exact number of goals scored by each team in each match is needed since drawn matches are common in this sport, and goal difference is the tie-breaking rule if two teams have the same score.
Football games are usually modelled such that the scores of both teams are described by a Poisson distribution, see, e.g.\ \citet{ChaterArrondelGayantLaslier2021} or \citet{DagaevRudyak2019}. Thus team $i$ scores $k$ goals against team $j$ with the probability
\begin{equation} \label{Poisson_dist}
P_k = \frac{\left( \lambda_{ij}^{(f)} \right)^k \exp\left( -\lambda_{ij}^{(f)} \right)}{k!},
\end{equation}
where $\lambda_{ij}^{(f)}$ shows the expected number of goals scored by team $i$ against team $j$, which depends on whether the match is played at home ($f = h$) or away ($f = a$).

We calculate parameter $\lambda_{ij}^{(f)}$ according to a function estimated on the basis of almost 30 thousand matches played by national football teams \citep{FootballRankings2020}. In particular, it is a quartic polynomial (a polynomial of degree four) of win expectancy with different equations for the home and away teams that contain a correction for unbalanced games.
The average expected number of goals for the home team $i$ is
\begin{equation} \label{Exp_goals_home}
\lambda_{ij}^{(h)} = 
\left\{ \begin{array}{ll}
-5.42301 \cdot W_{ij}^4 + 15.49728 \cdot W_{ij}^3 - \\
- 12.6499 \cdot W_{ij}^2 + 5.36198 \cdot W_{ij} + 0.22862 & \textrm{if } W_{ij} \leq 0.9 \\
231098.16153 \cdot (W_{ij}-0.9)^4 - 30953.10199 \cdot (W_{ij}-0.9)^3 + & \\
+ 1347.51495 \cdot (W_{ij}-0.9)^2 - 1.63074 (W_{ij}-0.9) + 2.54747 & \textrm{if } W_{ij} > 0.9
\end{array} \right.
\end{equation}
with $R^2 = 0.984$, whereas the average number of goals for the away team $j$ equals
\begin{equation} \label{Exp_goals_away}
\lambda_{ij}^{(a)} = 
\left\{ \begin{array}{ll}
90173.57949 \cdot (W_{ij} - 0.1)^4 + 10064.38612 \cdot (W_{ij} - 0.1)^3 + \\
+ 218.6628 \cdot (W_{ij} - 0.1)^2 - 11.06198 \cdot (W_{ij} - 0.1) + 2.28291 & \textrm{if } W_{ij} < 0.1 \\
-1.25010 \cdot W_{ij}^4 -  1.99984 \cdot W_{ij}^3 + & \\
+ 6.54946 \cdot W_{ij}^2 - 5.83979 \cdot W_{ij} + 2.80352 & \textrm{if } W_{ij} \geq 0.1
\end{array} \right.
\end{equation}
with $R^2 = 0.955$. The same model has been used recently to measure the probability of qualification for the 2018 FIFA World Cup \citep{Csato2022c}.

The final standing of the groups is determined in accordance with the criteria of the official regulation \citep[Article~20, Item~6]{FIFA2021}:
a) greatest number of points obtained in all group matches;
b) goal difference in all group matches;
c) greatest number of goals scored in all group matches.
Further tie-breaking rules are ignored for the sake of simplicity, that is, every remaining tie is decided randomly.

Besides the group rankings, the ranking of 2020/21 UEFA Nations League group winners determines the last two teams that enter the qualifying play-offs. However, the final positions of the top four Nations League group winners turn out only after the group stage draw in the qualifiers, hence they are ranked according to the principle in other leagues, see Table~\ref{Table1}.
In addition, at least 13 Nations League group winners may finish as group winners or runners-up in the European Qualifiers for the 2022 FIFA World Cup. The official regulation \citep{FIFA2020b} does not discuss this possibility, therefore the remaining vacancies in the play-offs are filled through the 2020/21 UEFA Nations League overall ranking.
These provisions are needed only to treat all possible cases and have at most marginal effect on the results.

\begin{table}[t!]
  \centering
  \caption{Strengths of the teams in the European Qualifiers for the 2022 FIFA World Cup}
  \label{Table2}
    \rowcolors{1}{gray!20}{}
    \begin{tabularx}{\textwidth}{m{4.5cm}CC C m{4.5cm}CC} \toprule \hiderowcolors
    Country & Rank & Elo &       & Country & Rank & Elo \\ \midrule
    \multicolumn{3}{c}{\textbf{Pot 1}} & \multicolumn{1}{c}{\textbf{}} & \multicolumn{3}{c}{\textbf{Pot 2}} \\ \bottomrule \showrowcolors
    \emph{Belgium} & 1     & 2111  &       & Switzerland & 11    & 1871 \\
    \emph{France} & 2     & 2092  &       & \emph{Wales} & 12    & 1829 \\
    England & 3     & 1962  &       & Poland & 13    & 1813 \\
    Portugal & 4     & 2038  &       & Sweden & 14    & 1821 \\
    \emph{Spain} & 5     & 2050  &       & \emph{Austria} & 15    & 1786 \\
    \emph{Italy} & 6     & 1998  &       & Ukraine & 16    & 1833 \\
    Croatia & 7     & 1860  &       & Serbia & 17    & 1789 \\
    Denmark & 8     & 1927  &       & Turkey & 18    & 1753 \\
    Germany & 9     & 1956  &       & Slovakia & 19    & 1664 \\
    Netherlands & 10    & 1994  &       & Romania & 20    & 1666 \\ \bottomrule
    \end{tabularx}

    \rowcolors{1}{}{gray!20}
    \begin{tabularx}{\textwidth}{m{4.5cm}CC C m{4.5cm}CC} \toprule
    \multicolumn{3}{c}{\textbf{Pot 3}} & \multicolumn{1}{c}{\textbf{}} & \multicolumn{3}{c}{\textbf{Pot 4}} \\ \bottomrule
    Russia & 21    & 1743  &       & Bosnia and Herzegovina & 31    & 1634 \\
    \emph{Hungary} & 22    & 1745  &       & \emph{Slovenia} & 32    & 1627 \\
    Republic of Ireland & 23    & 1658  &       & \emph{Montenegro} & 33    & 1531 \\
    \emph{Czech Republic} & 24    & 1761  &       & North Macedonia & 34    & 1570 \\
    Norway & 25    & 1763  &       & \emph{Albania} & 35    & 1556 \\
    Northern Ireland & 26    & 1586  &       & Bulgaria & 36    & 1505 \\
    Iceland & 27    & 1633  &       & Israel & 37    & 1592 \\
    Scotland & 28    & 1660  &       & Belarus & 38    & 1496 \\
    Greece & 29    & 1627  &       & Georgia & 39    & 1507 \\
    Finland & 30    & 1708  &       & Luxembourg & 40    & 1363 \\ \bottomrule
    \end{tabularx}  
   
\begin{threeparttable}
    \rowcolors{1}{}{gray!20}
    \begin{tabularx}{\textwidth}{m{4.5cm}CC C m{4.5cm}CC} \toprule
    \multicolumn{3}{c}{\textbf{Pot 5}} & \multicolumn{1}{c}{\textbf{}} & \multicolumn{3}{c}{\textbf{Pot 6}} \\ \bottomrule
    \emph{Armenia} & 41    & 1523  &       & Malta & 51    & 1215 \\
    Cyprus & 42    & 1326  &       & Moldova & 52    & 1227 \\
    \emph{Faroe Islands} & 43    & 1262  &       & Liechtenstein & 53    & 1115 \\
    Azerbaijan & 44    & 1383  &       & \emph{Gibraltar} & 54    & 1095 \\
    Estonia & 45    & 1383  &       & San Marino & 55    & 830 \\
    Kosovo & 46    & 1491  &       &       &       &  \\
    Kazakhstan & 47    & 1334  &       &       &       &  \\
    Lithuania & 48    & 1370  &       &       &       &  \\
    Latvia & 49    & 1261  &       &       &       &  \\
    Andorra & 50    & 1035  &       &       &       &  \\ \bottomrule
    \end{tabularx}
\begin{tablenotes} \footnotesize
\item
Group winners in the 2020/21 UEFA Nations League are written in \emph{italics}, see Table~\ref{Table1}.
\item
The column Rank shows the (internal) ranking according to the November 2020 FIFA World Ranking. Source: \url{https://www.fifa.com/fifa-world-ranking/ranking-table/men/rank/id13113/#UEFA}.
\item
The column Elo shows the World Football Elo Ratings as of 26 November 2020 (the day of the FIFA World Ranking). Source: \url{https://www.international-football.net/elo-ratings-table?year=2020&month=11&day=26&confed=UEFA}.
\end{tablenotes}
\end{threeparttable}
\end{table}

Table~\ref{Table2} presents the Elo ratings of the 55 UEFA member associations, which are ranked on the basis of the November 2020 FIFA World Ranking, underlying the draw of the qualification tournament.
As it has been discussed, these data are sufficient to simulate match outcomes due to formulas~\eqref{Elo_equation}, \eqref{Poisson_dist}, \eqref{Exp_goals_home}, and \eqref{Exp_goals_away}.
Table~\ref{Table3} presents the associated probabilities for a supposed match between Albania and Hungary, as well as between Belgium and San Marino, the strongest and the weakest teams in our study, respectively.

\begin{table}[t!]
  \centering
  \caption{Illustration: the probability of different match outcomes}
  \label{Table3}
    \rowcolors{1}{gray!20}{}
    \begin{tabularx}{\textwidth}{LL cccc} \toprule \hiderowcolors
    \multirow{2}[0]{*}{Home team} & \multirow{2}[0]{*}{Away team} & Win expectancy of & \multicolumn{3}{c}{The probability of} \\
          &       & the home team & Home win & Draw  & Away win \\ \bottomrule \showrowcolors
    Albania & Hungary & 0.3746 & 0.3132 & 0.2633 & 0.4235 \\
    Hungary & Albania & 0.8407 & 0.7225 & 0.1737 & 0.1037 \\
    Belgium & San Marino & 0.9996 & 0.9986 & 0.0012 & 0.0002 \\
    San Marino & Belgium & 0.0011 & 0.0041 & 0.0226 & 0.9734 \\ \bottomrule
    \end{tabularx}
\end{table}

First, the whole qualifying tournament is simulated to obtain the set of directly qualified teams (the group winners) and the set of teams progressing to the play-offs (the runners-up and the two participants chosen based on the Nations League ranking).
Second, it is checked whether any team would be better off by unilateral losing as follows. Assume that the team ranked $\tau_3$th among the Nations League group winners does not qualify for the play-offs but the teams ranked $\tau_1$th and $\tau_2$th ($\tau_1 < \tau_2 < \tau_3$) do qualify through the Nations League ranking, namely, without being group winners or runners-up in the qualifiers. The team ranked $\tau_3$th will be better off by losing if:
\begin{itemize}
\item
the team ranked $\tau_1$th among the Nations League group winners plays in its group, and the team ranked $\tau_1$th becomes a group winner or runner-up after the team ranked $\tau_3$th loses both of its matches against the team ranked $\tau_1$th; or
\item
the team ranked $\tau_2$th among the Nations League group winners plays in its group, and the team ranked $\tau_2$th becomes a group winner or runner-up after the team ranked $\tau_3$th loses both of its matches against the team ranked $\tau_2$th.
\end{itemize}
In the simulation, tanking of a team is achieved by adding 100 goals for its opponent in both matches.
If only the first condition (concerning the team ranked $\tau_1$th) holds, it is counted as an \emph{effective tanking} between the teams ranked $\tau_1$th and $\tau_3$th. If only the second (concerning the team ranked $\tau_2$th) holds, it means an \emph{effective tanking} between the teams ranked $\tau_2$th and $\tau_3$th. If both conditions are satisfied, the effective ranking counter is increased by half between the teams ranked $\tau_1$th and $\tau_3$th, as well as by half between the teams ranked $\tau_2$th and $\tau_3$th. Therefore, any situation in which the team ranked $\tau_3$th would be better off by unilateral losing is counted only once.

Theoretically, the next team ranked $\tau_4$th among the Nations League group winners that do not qualify for the play-offs might also have a small chance to benefit from tanking. This is possible if two of the teams ranked $\tau_1$th, $\tau_2$th, and $\tau_3$th become the group winner and runner-up by a strategic manipulation of the team ranked $\tau_4$th. Unsurprisingly, we have not found such an example in any simulation run since the conditions are strongly restrictive.
Finally, no team ranked lower than $\tau_4$th among the Nations League group winners can be better off by losing as there are at least four higher-ranked teams outside the top two in their groups, and at most two of them can achieve these positions after tanking. To conclude---similar to the concept of Nash equilibrium---the above procedure examines only unilateral losing strategies but it correctly identifies all possible cases of incentive incompatibility.

\begin{example} \label{Examp2}
Take the European Qualifiers for the 2022 FIFA World Cup and suppose that France (Group D), Belgium (Group E), Italy (Group C), and Spain (Group B) are group winners, whereas Austria (Group F), the Czech Republic (Group E), Slovenia (Group H), and Montenegro (Group G) are runners-up. If Hungary does not reach the top two positions in Group I, then Wales (Group E) and Hungary qualify for the play-offs as Nations League group winners, see Table~\ref{Table1}. With the notations above, Wales is the $\tau_1$th ranked team, Hungary is the $\tau_2$th, Albania (Group I) is the $\tau_3$th, and Armenia (Group J) is the $\tau_4$th. Wales and Albania play in different groups, thus Albania cannot tank against Wales. Since Hungary and Albania are assigned to the same group, it should be examined what happens if Hungary scores 100 additional goals in both of its matches against Albania. If Hungary does not become the group winner or runner-up after this change, there exists no effective tanking. On the other hand, if Hungary reaches the top two positions due to the modification, an effective tanking is booked in the relation of Hungary and Albania.
Armenia cannot qualify for the play-offs by unilateral losing because none of the three higher-ranked teams plays in its group.
\end{example}

A simulation run consists of the following steps:
\begin{enumerate}
\item
The groups of the European Qualifiers for the 2022 FIFA World Cup are drawn.
\item
The qualification matches are played; group rankings are established; group winners, runners-up, and teams advancing to the play-offs through the Nations League are determined.
\item
Unilateral losing by the teams ranked $\tau_3$th and $\tau_4$th among Nations League group winners is studied separately. Group rankings are established; group winners, runners-up, and teams advancing to the play-offs through the Nations League are determined.
\end{enumerate}

The error of a simulated probability $p$ is $\sqrt{p(1-p)/n}$ for sample size $n$.\footnote{~We are grateful to an anonymous reviewer for this important remark.}
As the probability of an effective tanking will turn out to be about 1.43\%, 10 million iterations is sufficient to get a reliable expected value: the error is 0.0038\% and the probability can be reduced below 0.013\% due to our recommendation. 

Teams can qualify for the FIFA World Cup directly or through the play-offs. Since exactly one team advances from each of the three play-off paths with four teams each, the aggregated probability of qualification is estimated by the probability of winning a group plus one-fourth of the probability that the team goes to the play-offs.

\section{Assessing and mitigating the problem of misaligned incentives} \label{Sec5}

\input{Figure2_effective_tanking_teams}

The first issue to be investigated is the probability of a situation where an effective tanking exists. It is about 1.4\%, which does not seem to be too worrying at first glance. On the other hand, Figure~\ref{Fig2} implies that in 0.8\% of all possible outcomes, Slovenia would be better off by losing against Hungary. Since the two teams have about a 10\% chance to play in the same group, the conditional probability of Slovenia tanking successfully against its neighbouring country is 8\%. The Czech Republic, Hungary, and Montenegro have also a relatively high chance to lose by exerting full effort. The organisers have clearly opted for a risky alternative by choosing this particular design: it would be scandalous and detrimental to the integrity of European football if any national team would miss qualification for the FIFA World Cup by playing honestly.

Figure~\ref{Fig2} suggests a straightforward way to reduce this unfairness. Note that incentive incompatibility cannot emerge in the relation of teams assigned to different groups. For example, the probability of effective tanking is zero for Austria against Wales, for Hungary against the Czech Republic, as well as for Montenegro against Slovenia, and for Albania against Slovenia or Montenegro. These teams are drawn from the same pot, thus they cannot play against each other in the qualifiers, which excludes any opportunity of tanking.

\begin{table}[t!]
  \centering
  \caption{Draw constraints (prohibited clashes) to mitigate incentive incompatibility}
  \label{Table4}
    \rowcolors{1}{}{gray!20}
    \begin{tabularx}{\textwidth}{LcLc} \toprule
    Higher-ranked team & Place in the draw & Lower-ranked team & Place in the draw \\ \bottomrule
    Wales & Pot 2 & Czech Republic & Pot 3 \\
    Austria & Pot 2 & Czech Republic & Pot 3 \\
    Wales & Pot 2 & Hungary & Pot 3 \\
    Austria & Pot 2 & Hungary & Pot 3 \\
    Wales & Pot 2 & Slovenia & Pot 4 \\
    Austria & Pot 2 & Slovenia & Pot 4 \\
    Czech Republic & Pot 3 & Slovenia & Pot 4 \\
    Hungary & Pot 3 & Slovenia & Pot 4 \\
    Austria & Pot 2 & Montenegro & Pot 4 \\
    Czech Republic & Pot 3 & Montenegro & Pot 4 \\
    Hungary & Pot 3 & Montenegro & Pot 4 \\
    Czech Republic & Pot 3 & Albania & Pot 4 \\
    Hungary & Pot 3 & Albania & Pot 4 \\
    Albania & Pot 4 & Armenia & Pot 5 \\ \bottomrule
    \end{tabularx}
\end{table}

Consequently, adding further draw constraints is a potent tool to avoid such an unwanted scenario. In particular, Table~\ref{Table4} presents 14 matches to prohibit in the tournament that we have chosen on the basis of Figure~\ref{Fig2} because they have a high vulnerability to effective tanking. This policy does not require any ``innovation'' from UEFA since there already exist similar prohibited country pairs in the draw due to political reasons, as well as draw conditions explained by competition-related issues \citep{UEFA2020c}. The new design with these additional restrictions is called \emph{constrained format}.

The choice of these additional restrictions is intuitive: Nations League group winners should play in different groups if they are close in the ranking of group winners (Table~\ref{Table1}) and face substantial uncertainty during the qualification. In other words, it is barely beneficial to prevent a match between Belgium and France (both teams finish almost always as group winners or runners-up in the qualifiers), between Austria and Gibraltar (the latter team has practically no chance to be better than the third-placed team of its group), or between Wales and Armenia (there are several Nations League group winners ranked between them, hence Armenia is unlikely to benefit from a loss against Wales). On the other hand, both the Czech Republic and Montenegro have a non-negligible probability to reach the first two positions and only two Nations League group winner are ranked between them, thus tanking by the latter team cannot be ruled out. Anyway, we have only used information available at the date of the draw, and the simulations can be carried out in a reasonable amount of time to determine the set of clashes that should be avoided.

Our proposal decreases the probability of effective tanking by two orders of magnitude, from more than 1.4\% to less than 0.013\%. Although misaligned incentives can still emerge for some teams, especially for Armenia (Figure~\ref{Fig2}), they do not pose a great threat---and can be entirely avoided by appropriate draw conditions.

\input{Figure3_probability_overview_both_formats}

The virtues of the constrained tournament design are reinforced by Figure~\ref{Fig3}, which conveys the key message of the paper. In the original format, three teams (Hungary, Slovenia, Montenegro) could increase their chance of qualification by more than 0.5\% if they choose a unilateral manipulation strategy. All teams that do not win their groups in the UEFA Nations League might lose due to misaligned incentives. On the other hand, adding a proper set of draw restrictions essentially eliminates any effect of incentive incompatibility.
The suggested policy reform is able to rectify a severe problem of the original tournament format, thus can greatly contribute to its spread.

\input{Figure4_constraints_effect}

Finally, it is worth considering how the alternative design affects the chance of playing in the FIFA World Cup. According to Figure~\ref{Fig4}, there is not much reason to worry in this respect: the changes in the probability of qualification remain below half percentage points, except for Slovenia and Montenegro, two Nations League group winners. They probably benefit from avoiding the Czech Republic and Hungary, the second and the third strongest team, respectively, in Pot 3 (see Table~\ref{Table2}). Furthermore, their gains from the additional draw restrictions can be easily justified because the tournament format depends on the results from the Nations League to a lesser degree than the UEFA Euro 2020 qualifying \citep[Chapter~6.4]{Csato2021a}.

The main features of the proposed solution can be summarised as follows:
\begin{itemize}
\item
It reduces the probability of incentive incompatibility by two orders of magnitude, therefore, it almost eliminates the danger of strategic manipulation.
\item
It marginally increases the chances of most Nations League group winners to play in the 2022 FIFA World Cup, thus it can intensify competition in the UEFA Nations League.
\item
Adding new draw constraints to the long list of current restrictions does not increase the complexity of tournament rules. Since, according to the official description of the draw procedure \citep{UEFA2020c}, these conditions \emph{aim to issue a schedule that is fair for the participating teams}, minimising the extent of incentive incompatibility is a persuasive argument for introducing further constraints.
In addition, protecting some Nations League group winners from competing against each other can be easily explained to the public.
\end{itemize}

\section{Conclusions} \label{Sec6}

This paper has come around the issue of how strategic manipulation can be prevented in a group-based qualification system with an exogenous ranking of the contestants used to provide a secondary way to qualify, where some teams might be interested in the success of other teams to create vacant slots to be filled. Although the theoretical background has been extensively discussed in the literature \citep{DagaevSonin2018, Csato2021a}, no attempt has ever been made to exceed the binary approach of incentive incompatibility by quantifying the probability of an effective tanking. To that end, the example of the European Qualifiers for the 2022 FIFA World Cup has been investigated as a case study. While strategy-proofness cannot be guaranteed without fundamentally changing the tournament design, adding a carefully chosen set of draw restrictions can considerably mitigate incentive incompatibility.

There are several avenues to continue our research. The strategic behaviour of the teams can be modelled by taking the situation before any particular match into account and considering a probabilistic assessment about the chance to qualify. One can test whether the assumed optimal choice is supported by evidence. A more sophisticated approach can be developed to predict the results of the games, even though that will probably not alter the main implications. Since many sports tournaments suffer from some problems around incentives, measuring the seriousness of such issues offers an obvious topic for future studies.
Hopefully, the present work will be only the first step in the quantification of incentive compatibility, which is indispensable to understand better the potential trade-offs between strategy-proofness and other desirable attributes of a tournament design.

Finally, one might legitimately ask whether our proposal has any chance to be implemented in practice. We are cautiously optimistic in this respect. First, adding further constraints to the draw procedure can be justified for the public by transparent competition-related reasons such as protecting certain Nations League group winners from playing against each other. Second, there exist some recent examples when the governing bodies in football have taken the results of academic researchers into account:
\begin{itemize}
\item
UEFA has modified the knockout bracket of the 2020 UEFA European Championship to balance group advantages based on the suggestion of \citet{Guyon2018a};
\item
Inspired by the criticism of \citet{Guyon2015a}, FIFA has reformed the draw of the 2018 FIFA World Cup;
\item
The recommendation of \citet{DuranGuajardoSaure2017} for a fairer schedule of the South American Qualifiers to the 2018 FIFA World Cup has been unanimously approved by the participants and is currently being used \citep{Alarconetal2017}.
\end{itemize}
Thus our study would nicely fit into the collection of academic works proposing rule change ideas that could yet be adopted and implemented by sports administrators \citep{LentenKendall2021}.
 
\section*{Acknowledgements}
\addcontentsline{toc}{section}{Acknowledgements}
\noindent
This paper could not have been written without \emph{my father} (also called \emph{L\'aszl\'o Csat\'o}), who has primarily coded the simulations in Python. \\
We are grateful to \emph{Dries Goossens}, \emph{D\'ora Gr\'eta Petr\'oczy}, and \emph{Zsombor Sz\'adoczki} for useful comments. \\
Three anonymous reviewers provided valuable remarks and suggestions on an earlier draft. \\
We are indebted to the \href{https://en.wikipedia.org/wiki/Wikipedia_community}{Wikipedia community} for summarising important details of the sports competition discussed in the paper. \\ 
The research was supported by the MTA Premium Postdoctoral Research Program grant PPD2019-9/2019.

\bibliographystyle{apalike}
\bibliography{All_references}

\end{document}

%% file: Figure2_effective_tanking_teams.tex
\begin{figure}[t!]
\centering

\begin{tikzpicture}
\begin{axis}[
name = axis1,
width = 0.93\textwidth, 
height = 1.2\textwidth,
xlabel style = {align=center, font=\small},
xlabel = {The number of effective tankings (out of $10^7$ runs, log scale)},
xmajorgrids = true,
xbar,
bar width = 4pt,
xmin = 0,
xmode = log,
scaled x ticks = false,
xticklabel style = {/pgf/number format/fixed,/pgf/number format/precision=5},
ytick style = {draw = none},
enlarge y limits = 0.04,
y dir = reverse,
symbolic y coords = {Top 4--WAL,Top 4--AUT,Top 4--CZE,WAL--CZE,AUT--CZE,Top 4--HUN,WAL--HUN,AUT--HUN,Top 4--SVN,WAL--SVN,AUT--SVN,CZE--SVN,HUN--SVN,Top 5--MNE,AUT--MNE,CZE--MNE,HUN--MNE,Top 6--ALB,CZE--ALB,HUN--ALB,Top 10--ARM,ALB--ARM,Top 12--GIB},
ytick = data,
legend entries = {Original format$\qquad$,Constrained format},
legend style = {at={(0.5,-0.08)},anchor = north,legend columns = 2}
]
\addplot [red, pattern color = red, pattern = grid, very thick] coordinates{
(14,Top 4--WAL)
(12,Top 4--AUT)
(37.5,Top 4--CZE)
(528.5,WAL--CZE)
(6101,AUT--CZE)
(15,Top 4--HUN)
(287,WAL--HUN)
(3608,AUT--HUN)
(49.5,Top 4--SVN)
(573.5,WAL--SVN)
(6669,AUT--SVN)
(28556.5,CZE--SVN)
(79219.5,HUN--SVN)
(84,Top 5--MNE)
(844,AUT--MNE)
(3781,CZE--MNE)
(9799,HUN--MNE)
(41,Top 6--ALB)
(168,CZE--ALB)
(411,HUN--ALB)
(408,Top 10--ARM)
(653,ALB--ARM)
(16,Top 12--GIB)
};

\addplot [blue, pattern color = blue, pattern = dots, very thick] coordinates{
(19,Top 4--WAL)
(14,Top 4--AUT)
(24,Top 4--CZE)
(0,WAL--CZE)
(0,AUT--CZE)
(15,Top 4--HUN)
(0,WAL--HUN)
(0,AUT--HUN)
(44,Top 4--SVN)
(0,WAL--SVN)
(0,AUT--SVN)
(0,CZE--SVN)
(0,HUN--SVN)
(205,Top 5--MNE)
(0,AUT--MNE)
(0,CZE--MNE)
(0,HUN--MNE)
(92,Top 6--ALB)
(0,CZE--ALB)
(0,HUN--ALB)
(791,Top 10--ARM)
(0,ALB--ARM)
(30,Top 12--GIB)
};
\end{axis}
\end{tikzpicture}

\caption{The threat of effective tanking for various country pairs \\ under the original and constrained tournament designs}
\label{Fig2}

\end{figure}
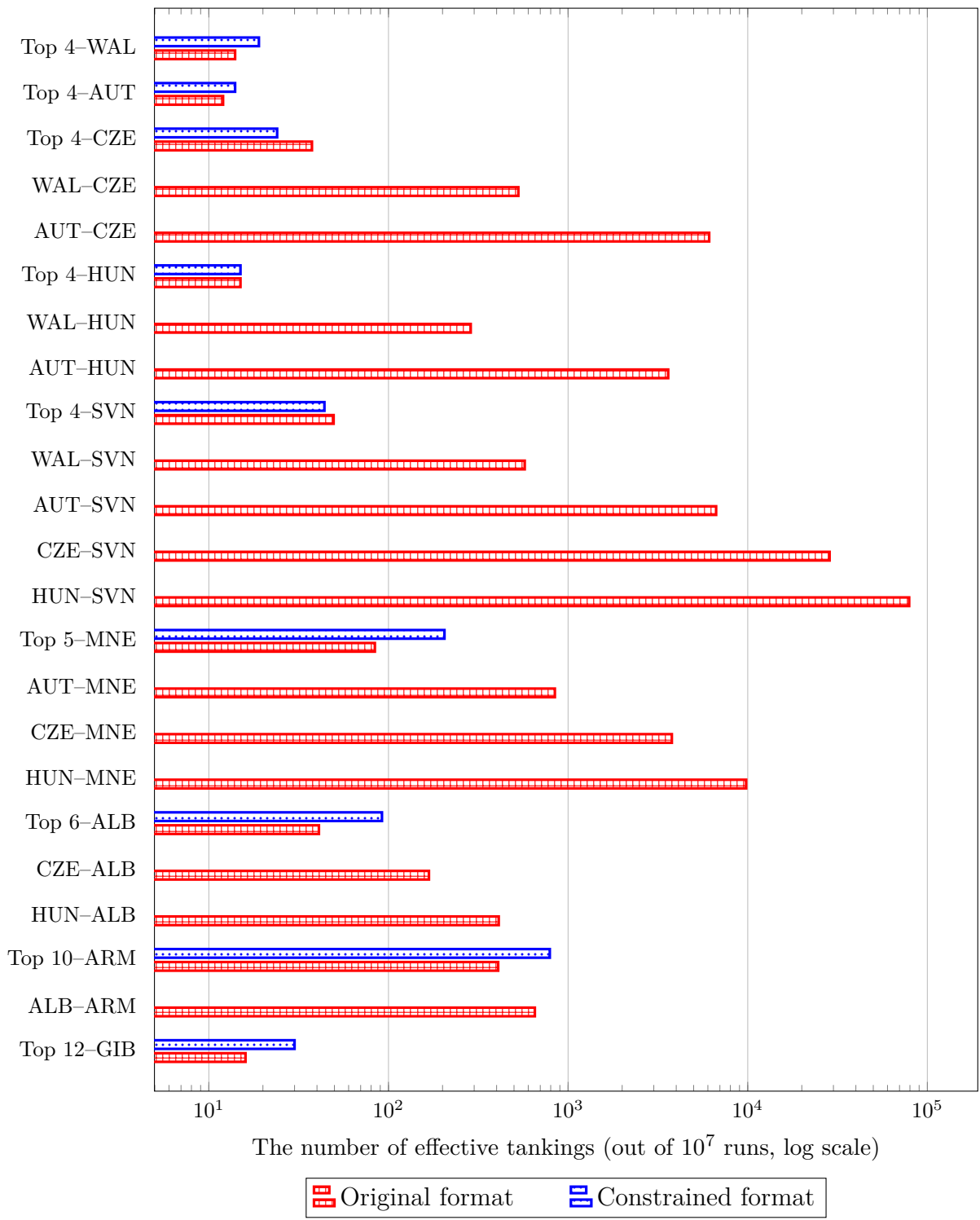


%% file: Figure3_probability_overview_both_formats.tex
\begin{figure}[t!]
\centering

\begin{tikzpicture}
\begin{axis}[width = 0.92\textwidth, 
height = 0.6\textwidth,
title style = {align = center,font = \small},
xmin = -0.02,
xmax = 1.02,
xmajorgrids,
ymajorgrids,
xlabel = {The probability of qualification in the original format},
xlabel style = {align = center,font =\small},
scaled y ticks = false,
y tick label style = {/pgf/number format/.cd,fixed,precision=4},
ylabel style = {font = \small, align=center},
ylabel = {Change in the probability of qualification \\ due to effective tanking},
legend entries = {{Nations League group winners, original format$\qquad \quad$},{Other teams, original format $\quad \;$},{Nations League group winners, constrained format$\quad$},{Other teams, constrained format}},
legend style = {at = {(0.43,-0.15)},anchor = north,legend columns = 2,font = \small}
]
\draw[thick](axis cs:\pgfkeysvalueof{/pgfplots/xmin},0)  -- (axis cs:\pgfkeysvalueof{/pgfplots/xmax},0);
\draw[thick,loosely dashed](0,0)  -- (axis cs:200*\pgfkeysvalueof{/pgfplots/ymax},\pgfkeysvalueof{/pgfplots/ymax}) node[pos=0.8, above left] {\footnotesize{$>0.5$\% gain}};
\draw[thick,densely dashed](0,0)  -- (axis cs:1000*\pgfkeysvalueof{/pgfplots/ymax},\pgfkeysvalueof{/pgfplots/ymax}) node[pos=0.3, above left] {\footnotesize{$>0.1$\% gain}};
\draw[thick,dotted](0,0)  -- (axis cs:-1000*\pgfkeysvalueof{/pgfplots/ymin},\pgfkeysvalueof{/pgfplots/ymin}) node[pos=0.68, below left] {\footnotesize{$>0.1$\% loss}};
\draw[thick,loosely dotted](0,0)  -- (axis cs:-400*\pgfkeysvalueof{/pgfplots/ymin},\pgfkeysvalueof{/pgfplots/ymin}) node[pos=0.68, below left] {\footnotesize{$>0.25$\% loss}};
\addplot[red,thick,only marks,mark=diamond*,mark size=3pt] coordinates {
(0.9540787,-6.39749999999939E-05)
(0.944596225,-7.61249999999825E-05)
(0.91782755,-9.84750000000423E-05)
(0.87352885,-0.000138300000000036)
(0.3848422,-5.26000000000137E-05)
(0.34329795,0.000166049999999973)
(0.2633593,0.00139242499999997)
(0.30246545,0.000618249999999987)
(0.1293239,0.00282417500000001)
(0.037604775,0.00034725)
(0.0199797,-2.09999999999794E-06)
(0.00965365,-7.89999999999992E-06)
(0.0000684,-2.49999999999993E-07)
(0.0000034,0.0000004)
};

\addplot[black,thick,only marks,mark=otimes*,mark size=2pt] coordinates {
(0.82745445,-0.000247024999999956)
(0.9080009,-0.000159849999999961)
(0.6551243,-0.000355424999999965)
(0.776961575,-0.000296674999999968)
(0.819199325,-0.000261750000000061)
(0.8662651,-0.000210550000000032)
(0.3902814,-0.000416699999999992)
(0.299990675,-0.000413799999999964)
(0.31194595,-0.000419325000000026)
(0.332158775,-0.000475350000000041)
(0.266868475,-0.000404650000000006)
(0.2190555,-0.000366824999999987)
(0.1223299,-0.000239275000000011)
(0.124078325,-0.000249675000000005)
(0.15960265,-3.81000000000131E-05)
(0.076939675,-2.26249999999983E-05)
(0.1787553,-3.73000000000179E-05)
(0.037708075,-1.18750000000015E-05)
(0.0607185,-0.000018650000000002)
(0.078485,-2.25999999999976E-05)
(0.0574148,-1.66750000000007E-05)
(0.118533375,-2.88999999999984E-05)
(0.04604355,-5.79750000000018E-05)
(0.0213292,-2.71750000000008E-05)
(0.008754725,-1.10500000000003E-05)
(0.02806145,-3.66250000000019E-05)
(0.00784485,-0.0000114)
(0.009065025,-1.18500000000007E-05)
(0.000884925,-1.09999999999997E-06)
(0.00028855,-1.12499999999999E-06)
(0.000901075,-3.10000000000002E-06)
(0.000956575,-3.14999999999994E-06)
(0.00618295,-2.63500000000005E-05)
(0.0003273,-1.37499999999997E-06)
(0.0007545,-2.50000000000001E-06)
(0.000074525,-4.24999999999997E-07)
(0.000000375,0)
(0.0000118,-2.50000000000006E-08)
(0.0000158,0)
(0.00000075,0)
(0,0)
};

\addplot[ForestGreen,very thick,only marks,mark=x,mark size=3pt] coordinates {
(0.9541069,-2.24999999964837E-07)
(0.944657725,-7.49999999882789E-08)
(0.9179356,3.37500000002766E-06)
(0.87366995,8.2499999998209E-07)
(0.386459225,2.42499999997259E-06)
(0.3450046,9.49999999999562E-07)
(0.263290275,6.75000000005532E-07)
(0.30352035,6.75000000005532E-07)
(0.141751775,2.37499999999891E-06)
(0.0496856,8.82499999999703E-06)
(0.02340055,2.12500000000213E-06)
(0.009733075,1.96499999999995E-05)
(0.000069075,0)
(0.0000116,0.00000075)
};

\addplot[blue,thick,only marks,mark=star,mark size=3pt] coordinates {
(0.827056975,-1.32499999994096E-06)
(0.907840675,-4.24999999970588E-07)
(0.654547675,-2.22500000002235E-06)
(0.776438775,-1.42499999999934E-06)
(0.819261625,-1.07500000001703E-06)
(0.86624445,-9.49999999999562E-07)
(0.3885811,-2.20000000000775E-06)
(0.2978966,-2.12500000001947E-06)
(0.30976415,-2.07499999999028E-06)
(0.329968825,-2.67500000000753E-06)
(0.2645924,-2.52500000003097E-06)
(0.216481975,-2.15000000000631E-06)
(0.119880825,-1.24999999999431E-06)
(0.1215797,-1.29999999999575E-06)
(0.157979025,-2.95000000000156E-06)
(0.076008025,-1.72500000000797E-06)
(0.177550675,-3.24999999998243E-06)
(0.0369417,-9.75000000000281E-07)
(0.059940725,-1.52500000000222E-06)
(0.0774769,-0.000002000000000002)
(0.056529475,-1.37500000000484E-06)
(0.11715495,-2.84999999999869E-06)
(0.0446692,-8.00000000002188E-07)
(0.020488175,-3.25000000002407E-07)
(0.008393975,-9.99999999994061E-08)
(0.0270602,-2.99999999998218E-07)
(0.0074799,-1.00000000000273E-07)
(0.008660625,-1.49999999999109E-07)
(0.00082175,0)
(0.0002859,0)
(0.000885975,0)
(0.000934125,-2.499999999996E-08)
(0.006146275,-1.49999999999977E-07)
(0.0003173,-2.50000000000142E-08)
(0.0007456,0)
(0.000070725,0)
(0.000000175,0)
(0.0000102,0)
(0.000015625,0)
(0.00000075,0)
(0,0)
};

\node[pin={[ultra thick] 360:{\textcolor{black}{\footnotesize{Austria}}}}] at (0.34329795,0.000166049999999973) {};

\node[pin={[ultra thick] 360:{\textcolor{black}{\footnotesize{Hungary}}}}] at (0.2633593,0.00139242499999997) {};

\node[pin={[ultra thick] 360:{\textcolor{black}{\footnotesize{Czech Republic}}}}] at (0.30246545,0.000618249999999987) {};

\node[pin={[ultra thick] 360:{\textcolor{black}{\footnotesize{Slovenia}}}}] at (0.1293239,0.00282417500000001) {};

\node[pin={[ultra thick] 360:{\textcolor{black}{\footnotesize{Montenegro}}}}] at (0.037604775,0.00034725) {};
\end{axis}
\end{tikzpicture}

\captionsetup{justification=centering}
\caption{The role of incentive incompatibility \\ in the original and constrained tournament designs} 
\label{Fig3}

\end{figure}
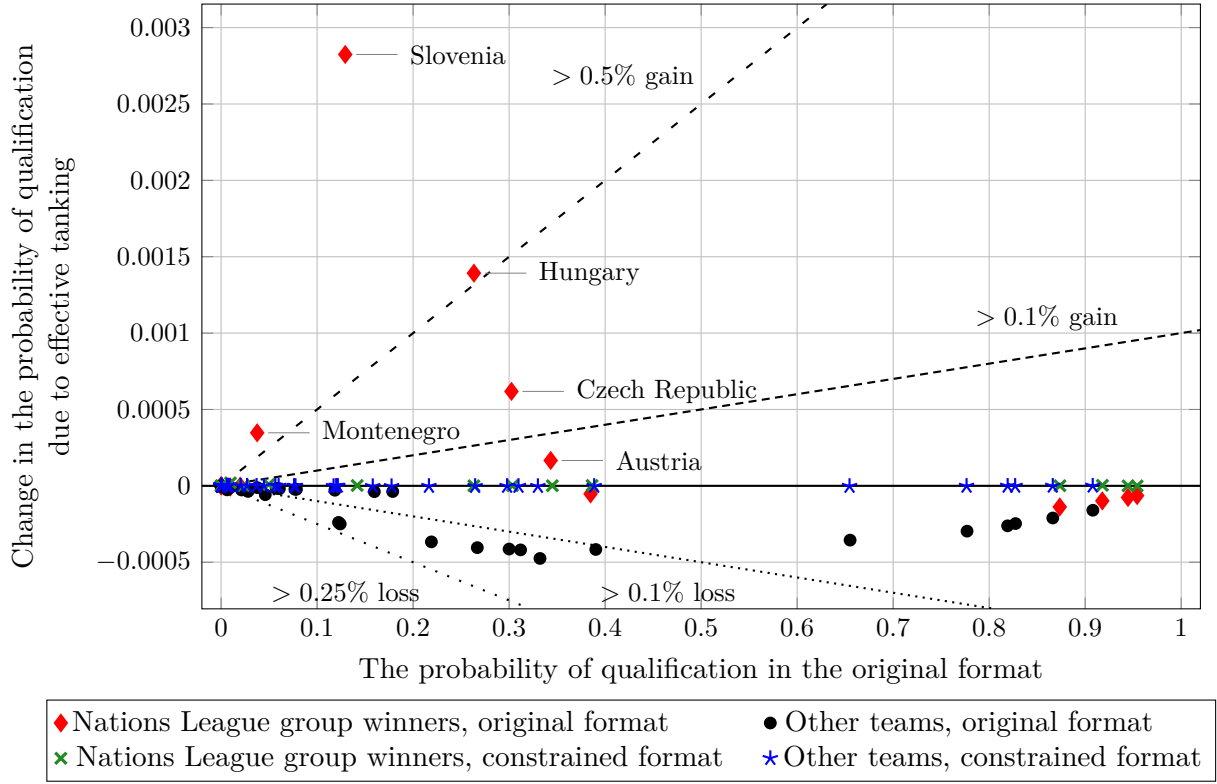


%% file: Figure4_constraints_effect.tex
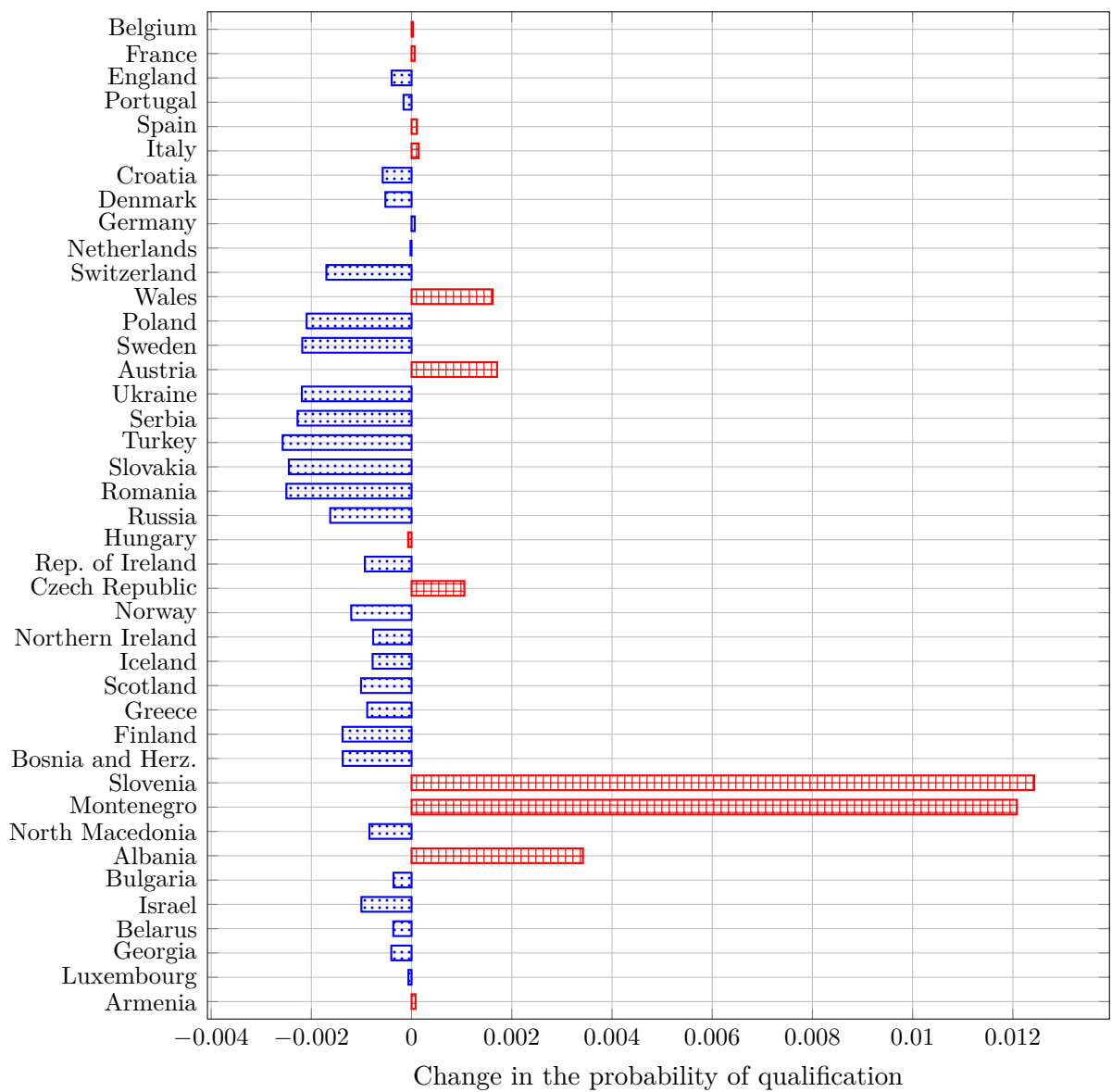
\begin{figure}[t!]
\centering

\begin{tikzpicture}
\begin{axis}[width = 0.9\textwidth, 
height = \textwidth,
xmajorgrids,
ymajorgrids,
xbar stacked,
bar width = 6pt,
scaled x ticks = false,
x tick label style = {/pgf/number format/.cd,fixed,precision=4},
xlabel = {Change in the probability of qualification},
xlabel style = {font = \small},
symbolic y coords = {Belgium, France, England, Portugal, Spain, Italy, Croatia, Denmark, Germany, Netherlands, Switzerland, Wales, Poland, Sweden, Austria, Ukraine, Serbia, Turkey, Slovakia, Romania, Russia, Hungary, Rep.\ of Ireland, Czech Republic, Norway, Northern Ireland, Iceland, Scotland, Greece, Finland, Bosnia and Herz., Slovenia, Montenegro, North Macedonia, Albania, Bulgaria, Israel, Belarus, Georgia, Luxembourg, Armenia},
ytick = data,
ylabel style = {font = \small},
y dir = reverse,
enlarge y limits = {abs = 0.25cm},
]
\addplot[blue,pattern = dots,pattern color = blue,thick] coordinates {
(0,Belgium)
(0,France)
(-0.000397475,England)
(-0.000160225,Portugal)
(0,Spain)
(0,Italy)
(-0.000576625,Croatia)
(-0.0005228,Denmark)
(0.0000623,Germany)
(-0.00002065,Netherlands)
(-0.0017003,Switzerland)
(0,Wales)
(-0.002094075,Poland)
(-0.0021818,Sweden)
(0,Austria)
(-0.00218995,Ukraine)
(-0.002276075,Serbia)
(-0.002573525,Turkey)
(-0.002449075,Slovakia)
(-0.002498625,Romania)
(-0.001623625,Russia)
(0,Hungary)
(-0.00093165,Rep.\ of Ireland)
(0,Czech Republic)
(-0.001204625,Norway)
(-0.000766375,Northern Ireland)
(-0.000777775,Iceland)
(-0.0010081,Scotland)
(-0.000885325,Greece)
(-0.001378425,Finland)
(-0.00137435,Bosnia and Herz.)
(0,Slovenia)
(0,Montenegro)
(-0.000841025,North Macedonia)
(0,Albania)
(-0.00036075,Bulgaria)
(-0.00100125,Israel)
(-0.00036495,Belarus)
(-0.0004044,Georgia)
(-0.000063175,Luxembourg)
(0,Armenia)
};

\addplot[red,pattern = grid,pattern color = red,thick] coordinates {
(0.0000282,Belgium)
(0.0000615,France)
(0,England)
(0,Portugal)
(0.00010805,Spain)
(0.0001411,Italy)
(0,Croatia)
(0,Denmark)
(0,Germany)
(0,Netherlands)
(0,Switzerland)
(0.001617025,Wales)
(0,Poland)
(0,Sweden)
(0.00170665,Austria)
(0,Ukraine)
(0,Serbia)
(0,Turkey)
(0,Slovakia)
(0,Romania)
(0,Russia)
(-0.000069025,Hungary)
(0,Rep.\ of Ireland)
(0.0010549,Czech Republic)
(0.012427875,Slovenia)
(0.012080825,Montenegro)
(0.00342085,Albania)
(0.000079425,Armenia)
};
\end{axis}
\end{tikzpicture}

\captionsetup{justification=centering}
\caption{The effect of additional draw constraints \\ \vspace{-0.25cm}
\begin{flushleft}
\footnotesize{\emph{Note}: Group winners in the 2020/21 UEFA Nations League are indicated by red bars with red grid. \\
Only the associations drawn from Pots 1--4 and Armenia are shown.}
\end{flushleft}}
\label{Fig4}

\end{figure}
